\documentclass[english,aps,prd,floats,floatfix,nofootinbib,preprintnumbers]{revtex4}
\usepackage{natbib}
\usepackage{times}
\usepackage{float}
\usepackage{amsmath}
\usepackage{graphicx}
\usepackage{amssymb}
\usepackage{slashed}
\usepackage[usenames]{xcolor}
\usepackage{soul} 
\usepackage{url}
\usepackage[colorlinks, hyperindex]{hyperref}
	\hypersetup
	{
		colorlinks, %
		citecolor=blue, %
		linkcolor=red, %
		urlcolor=blue, %
	}
\usepackage{doi}

\newcommand{\eL}{\epsilon_L}
\newcommand{\eR}{\epsilon_R}
\newcommand{\eS}{\epsilon_S}
\newcommand{\eP}{\epsilon_P}
\newcommand{\eT}{\epsilon_T}

\begin{document}
\preprint{CERN-TH-2016-230}
 \title{The lifetime of the $B_c^-$ meson and the anomalies in $B\to D^{(*)}\tau\nu$}
\author{Rodrigo Alonso$^1$, Benjam\'in Grinstein$^2$ and Jorge Martin Camalich$^1$}
\affiliation{$^1$CERN, Theoretical Physics Department, Geneva, Switzerland\\
$^2$Dept. Physics, University of California, San Diego, 9500 Gilman Drive, La Jolla, CA 92093-0319, USA}

\begin{abstract}
We investigate a new constraint on new-physics interpretations of the anomalies observed in $B\to D^{(*)}\tau\nu$ decays
making use of the lifetime of the $B_c^-$ meson. A constraint is obtained by 
demanding that the rate for  $B_c^-\to\tau^-\bar\nu$ does not exceed the fraction of the total width that is allowed by the calculation of the lifetime in the standard model. This leads to a very strong bound on new-physics scenarios involving scalar operators since they lift the slight, but not negligible, chiral suppression of the $B_c^-\to\tau^-\bar\nu$ amplitude in the standard model. The new constraint renders a scalar interpretation of the enhancement measured in $R_{D^*}$ implausible, including explanations implementing extra Higgs doublets or certain classes of leptoquarks. We also discuss the complementarity of $R_{D^{(*)}}$ and a measurement of the longitudinal polarization of the $\tau$ in the $B\to D^*\tau\nu$ decay in light of our findings. 
\end{abstract}

\maketitle

\section{Introduction}

The last decade has witnessed a very rapid development of the experimental measurement and theoretical understanding of many $B$-meson transitions. Although no observations of new particles have been reported at the LHC, some discrepancies with the predictions of the Standard Model (SM) have started to appear in semileptonic $B$ decays~\cite{Aaij:2014ora,Aaij:2015oid,Wehle:2016gfb,Lees:2012xj,Lees:2013uzd,Huschle:2015rga,Sato:2016svk,Abdesselam:2016xqt}. For instance, measurements of the charged-current $b\to c\tau\nu$ transitions turn out to be enhanced with respect to the SM. These have been measured through the lepton-universality ratios $R_{D^{(*)}} = \Gamma(B\to D^{(*)} \tau \nu)/\Gamma(B\to D^{(*)} \ell \nu)$ (henceforth $\ell$ stands for the muon or the electron) by two different experiments for the $B\to D \tau \nu$ channel, Babar~\cite{Lees:2012xj,Lees:2013uzd} and Belle~\cite{Huschle:2015rga,Sato:2016svk,Abdesselam:2016xqt}, and also by LHCb~\cite{Aaij:2015yra} for the $B\to D^{*} \tau \nu$ one. On the other hand, the theoretical predictions are very accurate as they only rely on parametrizations of the experimental spectra of the $B\to D^{(*)} \ell \nu$ decays with form factors described within the heavy-quark expansion and incorporating constraints from unitarity~\cite{deRafael:1993ib,Boyd:1995sq,Boyd:1997kz,Caprini:1997mu}.
Moreover, calculations of the relevant form factors beyond the zero-recoil limit in lattice QCD have started to appear~\cite{Lattice:2015rga,Na:2015kha,Du:2015tda}. The global significance of these anomalies currently stands at $\sim 4\sigma$ and they have been addressed in many different models of new physics (NP)~\cite{Tanaka:2012nw,Fajfer:2012vx,Becirevic:2012jf,Datta:2012qk,Duraisamy:2013kcw,Biancofiore:2013ki,Bhattacharya:2015ida,Alok:2016qyh,Feruglio:2016gvd,Faroughy:2016osc,Bardhan:2016uhr,Hou:1992sy,Tanaka:1994ay,Kiers:1997zt,Chen:2006nua,Crivellin:2012ye,Celis:2012dk,Crivellin:2013wna,Cline:2015lqp,Kim:2015zla,Crivellin:2015hha,Wang:2016ggf,Ko:2012sv,Sakaki:2013bfa,Sakaki:2014sea,Alonso:2015sja,Freytsis:2015qca,Barbieri:2015yvd,Calibbi:2015kma,Fajfer:2015ycq,Bauer:2015knc,Hati:2015awg,Sahoo:2015pzk,Zhu:2016xdg,Dumont:2016xpj,Das:2016vkr,Li:2016vvp,Bhattacharya:2016mcc,Becirevic:2016yqi,Sahoo:2016pet}.

An efficient strategy to analyse the possible NP scenarios is using a bottom-up approach in effective field theory (EFT), which starts with the most-general effective Lagrangian valid at the energy scale characteristic of the physical decay process, $\mu\sim m_B$:
\begin{multline}
\label{eq:leff1} 
{\cal L}_{\rm eff} 
=
- \frac{4G_F V_{cb}}{\sqrt{2}} \,
\Bigg[
\Big(1 + \epsilon_L \Big) \bar{\tau}  \gamma_\mu  P_L \nu_{\tau} \cdot \bar{c}   \gamma^\mu P_L b+\eR \bar{\tau}  \gamma_\mu  P_L   \nu_{\tau} \cdot \bar{c}   \gamma^\mu P_R b\\
+  \epsilon_T   
\,   \bar{\tau}   \sigma_{\mu \nu} P_L \nu_{\tau}    \cdot  \bar{c}   \sigma^{\mu \nu} P_L b+ \epsilon_{S_L}\bar{\tau}  P_L \nu_{\tau}\cdot \bar{c}P_L b+\epsilon_{S_R}\bar{\tau}  P_L\nu_{\tau}\cdot\bar{c}P_R b
\Bigg]+{\rm h.c.}
\end{multline}
In the construction of this low-energy EFT (LEEFT) we have used SM particles only. Furthermore, if there is a mass gap between the NP and electroweak-symmetry breaking (EWSB) scales, as the current empirical evidence seems to suggest, then one can express the low-energy operators in terms of $SU(2)_L\times U(1)_Y$-invariant ones~\cite{Buchmuller:1985jz,Cirigliano:2009wk,Grzadkowski:2010es}. In the context of this EFT of the SM (SMEFT), it follows that, at leading order in the decoupling limit ($\sim v^2/\Lambda^2$), $\eR$ is independent of the
charged-lepton flavor~\cite{Bernard:2006gy}  and it  cannot explain lepton-universality violation in $R_{D^{(*)}}$~\cite{Alonso:2015sja}.\footnote{See Ref.~\cite{Cata:2015lta} for a discussion of the consequences in the EFT arising with a nonlinear realization of EWSB.}

Nonetheless, a model that has been very popular in the interpretation of these anomalies
is the two-Higgs-doublet model (2HDM)~\cite{Branco:2011iw} which adds an extra Higgs
doublet to the SM and induces effective scalar couplings $\epsilon_{S_{L,R}}$
mediated by a charged Higgs, $H^\pm$. Focusing on implementations with discrete symmetries to avoid flavour-changing neutral currents, the resulting LEEFT of Eq.~(\ref{eq:leff1}) satisfies the minimal-flavor-violation and minimal-lepton-flavor-violation hypothesis, and as a result $\epsilon_{L,R}$ are lepton-flavor invariant up to order $(m_\tau/v)^2$, while $\epsilon_{S_{L,R}}^l\propto m_l/v$ break lepton-flavor at leading order in the lepton mass. This naturally explains the characteristic flavor structure required from the interaction, coupling selectively to the heavy leptons
to explain $R_{D^{(*)}}$ and secluded from the very stringent tests of lepton universality 
involving light quarks done with pion, kaon and nuclear (neutron) $\beta$ decays~\cite{Gonzalez-Alonso:2016etj}.  
In fact, collider and low-energy flavor data still allow for all the Higgs bosons emerging in the 2HDM to have masses at the EWSB scale~\cite{Akeroyd:2016ymd}. If one only adds this light-Higgs
doublet to the particle content of the SM, and still assumes there is a mass gap
with respect to any other NP scale, then the resulting 2HDM EFT~\cite{Crivellin:2016ihg}
still cannot produce a non-lepton-universal coupling $\eR$ at leading order in the
decoupling limit.

Another NP solution to the anomalies that have been discussed
extensively in the literature invoke the presence of leptoquarks. As colored particles,
the leptoquarks should be produced copiously in $pp$ collisions at the LHC, which leads to
stringent lower bounds on their mass currently reaching the TeV scale. This justifies
studying their effects at lower energies by matching first the
corresponding model to the SMEFT~\cite{Alonso:2015sja}.  

Thus, within these assumptions in the EFT framework (encompassing both the SMEFT and the 2HDM EFT) one has a model-independent parametrization of any possible NP contribution in the amplitudes and observables in terms of  the four Wilson coefficients $\epsilon_L$, $\epsilon_{S_L}$, $\epsilon_{S_R}$ and $\epsilon_T$. In order to narrow down the \textit{shape} of the putative NP effect and discriminate among different models, new observables are necessary which have different sensitivities 
to the various $\epsilon_i$. Examples that have been proposed in the literature include the polarization observables related to the $\tau$ in 
the decay~\cite{Bullock:1992yt,Tanaka:1994ay,Tanaka:2010se,Abdesselam:2016xqt}, the kinematic distributions~\cite{Lees:2013uzd,Freytsis:2015qca},
in particular, in terms of the final observable decay products~\cite{Nierste:2008qe,Sakaki:2012ft,Hagiwara:2014tsa,Bordone:2016tex,Alonso:2016gym,Ligeti:2016npd} or all the very same observables appearing in the equivalent decay modes of other bottom hadrons~\cite{CiezarekLHCb}, such as $B_s$~\cite{Bhol:2014jta}, 
$B_c$~\cite{Lytle:2016ixw} and $\Lambda_b$~\cite{Detmold:2015aaa}. 

In this work, we investigate a powerful constraint on NP that can be derived from the lifetime of the $B_c$ meson~\cite{Li:2016vvp}. This is obtained 
by demanding that the total rate of the $B_c^-\to\tau^-\bar\nu$ decay does not exceed the fraction of the total width that is allowed
by the calculation of the lifetime in the SM. Since the chirally-flipped operators lift the chiral suppression that the amplitude
receives in the SM, one obtains a very strong bound on the scalar operators. We discuss how this new observation puts considerable stress on
\textit{any} ``scalar interpretation'' of $R_{D^*}$, such as those produced by the 2HDM and certain leptoquark models. Finally, we discuss the 
interplay also between $R_D^{(*)}$ and the longitudinal polarization of the $\tau$ in the $B\to D^*\tau\nu$ decay in light of our findings and the recent
measurement of Belle~\cite{Abdesselam:2016xqt}.\footnote{For all the calculations related to the $B\to D^{(*)}\tau\nu$ decay modes we follow~\cite{Alonso:2016gym}.}  

\section{Interplay between the $B_c$ lifetime and $R_{D^*}$}

Parity conservation of the strong interactions implies that only the hadronic form factors for 
the pseudoscalar (scalar) combination $\eP=\epsilon_{S_R}-\epsilon_{S_L}$ ($\eS=\epsilon_{S_R}+\epsilon_{S_L}$) can contribute to $R_{D^*}$ ($R_D$) 
mode (see for instance~\cite{Manohar:2000dt}). This immediately links the $R_{D^*}$ 
anomaly to the tauonic decay of the $B_c^-$, $B_c^-\to\tau^-\bar \nu$, for the latter 
also receives a contribution from $\epsilon_P$ with a slight (but non-negligible) enhancement with respect to the SM produced by the chirally-flipped
nature of the scalar operators. Namely, from the effective Lagrangian in Eq.~(\ref{eq:leff1}) one obtains~\cite{Gonzalez-Alonso:2016etj}:
\begin{eqnarray}
{\rm Br}(B_c^-\to\tau\bar\nu_\tau)=\tau_{B_c^-}\,\frac{m_{B_c}m_\tau^2 f_{B_c}^2G_F^2 |V_{cb}|^2}{8\pi}\left(1-\frac{m_\tau^2}{m_{B_c}^2}\right)^2\,
\left|1+\epsilon_L+\frac{m_{B_c}^2}{m_\tau(m_b+m_c)}\eP\right|^2,
\label{eq:Bctotaunu}
\end{eqnarray}
where $f_{B_c}=434(15)$ MeV~\cite{Colquhoun:2015oha} is the $B_c$ decay constant and $m_b$ and $m_c$ are the quark masses in the $\overline{MS}$ 
evaluated at $\mu=2$ GeV (and so is $\eP$). Note that $\epsilon_T$ does not enter the decay as opposed to $\epsilon_R$ which does but its not included since we study lepton-universality violation.
Therefore, any experimental measurement or bound on ${\rm Br}(B_c^-\to\tau\bar\nu_\tau)$
can help constraining $\eP$, and to a lesser extent, $\eL$.

The $B_c$ enjoys high production rates at the LHC~\cite{Gouz:2002kk} and precise
measurements of the lifetime~\cite{Aaij:2014gka,Agashe:2014kda} or the branching ratios of many of its nonleptonic exclusive decay modes
are now available~\cite{Anderlini:2014dha,Dey:2016qck}. However, a measurement of the tauonic decay branching fraction, giving \textit{direct} access to the NP 
operators in Eq.~(\ref{eq:Bctotaunu}), seems to be out of reach~\cite{Gouz:2002kk}. An \textit{indirect} constraint on NP can be extracted, though, 
from the experimentally measured $B_c$ lifetime itself~\cite{Li:2016vvp}, whose PDG value is~\cite{Agashe:2014kda},
\begin{equation}
\tau_{B_c}=0.507(8)~\text{ ps}, \label{eq:Bc_lifetime_expt}
\end{equation}
and the fact that it should be mainly accounted for by $b$ and $c$ decays in the $B_c$ meson~\cite{Gershtein:1994jw,Bigi:1995fs}.

Indeed, the lifetime of the $B_c$ can be calculated in the SM computing the inclusive $\sum b\to c$, $\sum \bar c\to \bar s$ and annihilation decay rates 
using an operator product expansion (OPE) matched to non-relativistic QCD~\cite{Bigi:1995fs,Beneke:1996xe,Chang:2000ac}, which lead to results 
consistent with the experimental measurement~\cite{Beneke:1996xe}, 
\begin{equation}
\tau_{B_c}^{\rm OPE}=0.52^{+0.18}_{-0.12}~\text{ ps}.\label{eq:Bc_lifetime_OPE}
\end{equation}
For the central value, the width of the $B_c$, $\Gamma_{B_c}^{\rm OPE}=1/\tau_{B_c}^{\rm OPE}$, is distributed among modes induced by the partonic 
transitions $\bar c\to \bar s \bar u d$ (47\%), $\bar c\to \bar s \ell \bar\nu$ (17\%),  $b\to c \bar u d$ (16\%), 
$b\to c \bar \nu \ell$  (8\%) and $b\to c \bar c s$ (7\%). Hence, according to this calculation, only $\lesssim5\%$ of the measured experimental width, 
$\Gamma_{B_c}=1/\tau_{B_c}$, can be explained by (semi)tauonic modes, including those NP effects that would explain the $R_{D^{(*)}}$ anomalies. 
Similar conclusions are derived from the other OPE 
calculations~\cite{Bigi:1995fs,Chang:2000ac} as well as from those obtained using potential models~\cite{Gershtein:1994jw} or QCD sum rules~\cite{Kiselev:2000pp} 
(see Ref.~\cite{Gouz:2002kk} for a comparison among the different predictions). This constraint can be relaxed up to a 
$\lesssim30\%$ of $\Gamma_{B_c}$ if the longer lifetime given by the upper limit in Eq.~(\ref{eq:Bc_lifetime_OPE}) is taken as an input for the SM calculation
in the OPE.

The discussion above is immaterial for a NP scenario involving the left-handed contribution $\eL$, which explains $R_{D^{(*)}}$ by an overall
$\sim15\%$ enhancement of the SM at the amplitude level (see below) and modifies the corresponding branching fraction for $B_c^-\to \tau^-\bar \nu$  ($\sim 2\%$) 
well below these limits. On the other hand, the $B_c$ lifetime becomes a very severe constraint for scenarios aiming to explain $R_{D^*}$ with pseudoscalar
contributions. Indeed, fitting the average of the experimental determinations~\cite{Amhis:2014hma},
\begin{equation}
R_{D^*}^{\rm expt}=0.316(16)(10),\label{eq:RDs_expt}
\end{equation}
to $\eP$ we find the following $1\sigma$ range,
\begin{equation}
\eP=1.48(34),\label{eq:eP_RDs}
\end{equation}
which is consistent with the fit in the very same scenario reported in~\cite{Bardhan:2016uhr} and where we have neglected another solution which
interferes destructively with the SM, $\eP\simeq-4.3$. The value in Eq.~(\ref{eq:eP_RDs}), 
together with the enhancement factor $m_{B_c}^2/(m_\tau(m_b+m_c))\simeq 4$, leads to a boost of the $B_c^-\to \tau^-\bar \nu$ branching fraction 
by a factor of no less than $\sim25$, that not only exceeds the limits discussed above but can also be larger than the experimental width of the $B_c$. 

\begin{figure}[h]
\begin{tabular}{cc}
\includegraphics[width=74mm]{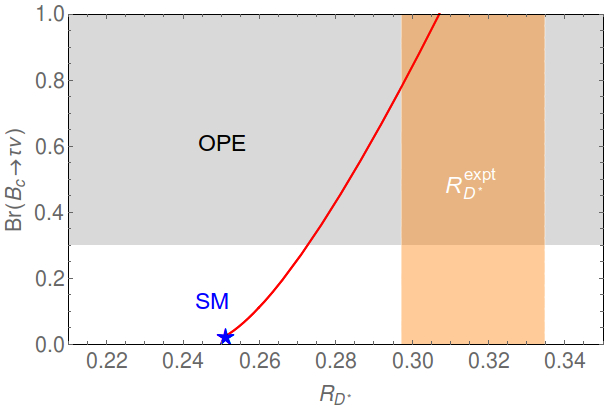}\hspace{1cm} &\includegraphics[width=78mm]{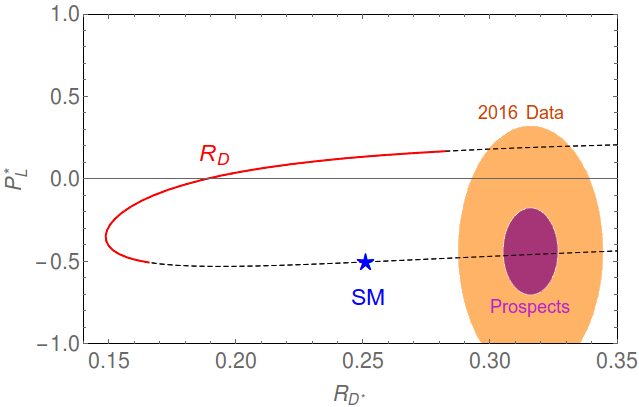}\\
\end{tabular}
\caption{\textit{Left-panel:} Correlation between $R_{D^*}$ and ${\rm Br}(B_c\to\tau\nu)$ for a pseudoscalar NP interaction (red line). The shaded areas
are the 1$\sigma$-band corresponding to the measurement of $R_{D^*}$ (vertical orange) and to the bound on the NP contribution to the 
lifetime of the $B_c$ assuming that the SM accounts for the 70\% of it (gray horizontal). \textit{Right-panel:} The (black) dashed line represents the parametric 
$\eT$-dependence of $R_{D^*}$ and $P_L^*$. The overlaid (red) solid line corresponds to values of $\eT$ for which $R_D$ is consistent with the experimental measurement at 
$1\sigma$. The shaded areas are the 1$\sigma$-bands corresponding to current data set (orange) 
and \textit{na\"ive} experimental prospects discussed in the main text (purple).
\label{fig:BctaunuRDs}}
\end{figure}

This is illustrated in Fig.~\ref{fig:BctaunuRDs} where we plot the correlation between $R_{D^*}$ and ${\rm Br}(B_c\to\tau\nu)$
in the presence of a pseudoscalar operator. The maximum value of $R_{D^*}$ that can be achieved without violating $\text{Br}(B_c\to\tau\nu)\lesssim 30\%$
is $R_{D^*}\simeq0.27$, corresponding to the value $\eP^{\rm max}=0.61$, and describing the experimental value for $R_{D^*}$ in Eq.~(\ref{eq:RDs_expt})
would saturate (or even exceed) the experimental width of the $B_c$ with the rate of the tauonic decay. 
The conclusion is that the $B_c$ lifetime makes highly implausible \textit{any} explanation of $R_{D^*}$ induced by $\eP$. 

In fact, this type of scenario encompasses models of NP that have been popular in the interpretation of the $R_{D^{(*)}}$, such as the 2HDM~\cite{Hou:1992sy,Tanaka:1994ay,Kiers:1997zt,Chen:2006nua,Crivellin:2012ye,Celis:2012dk,Crivellin:2013wna,Cline:2015lqp,Kim:2015zla,Crivellin:2015hha,Wang:2016ggf}. In light of the constraint from the $B_c$ lifetime, any of the 2HDM interpretations of $R_{D^*}$ is ruled out, in particular those  realizations of the model based on the so-called type III, with general Yukawa couplings to the different fermions~\cite{Branco:2011iw}, which can explain simultaneously $R_D$ and $R_{D^{*}}$~\cite{Crivellin:2012ye,Crivellin:2013wna,Cline:2015lqp,Kim:2015zla,Crivellin:2015hha,Wang:2016ggf} unlike the 2HDM tested by BaBar~\cite{Lees:2012xj,Lees:2013uzd} and Belle~\cite{Huschle:2015rga}. Model-building possibilities invoking leptoquarks are also subject to the bound from the lifetime of the $B_c$ meson if they generate sizable scalar operators~(see for instance ref.~\cite{Li:2016vvp} for a first discussion of this).

\section{Interplay between $P_L^*$ and $R_{D^{(*)}}$}

The analysis of the $B_c$ lifetime leaves only $\eL$ and $\eT$ as possible mechanisms to explain the $R_{D^*}$ anomaly. As discussed
succinctly above, the left-handed scenario leads to a \textit{universal} enhancement with respect to the SM of \textit{all} the $b\to c\tau\nu$ 
decay rates by the very same $\sim30\%$ that is required to simultaneously explain $R_D$ and $R_D^{*}$. Moreover, observables of a given semitauonic decay, 
normalized by the corresponding total rate of the \textit{same} decay, must be insensitive to $\eL$ because the enhancements cancel in the ratio.
The tensor (and scalar) contribution,
on the other hand, has a different structure that is manifest not only in the total rate but also in the kinematic distributions of 
the decay~\cite{Tanaka:2012nw,Biancofiore:2013ki,Bhattacharya:2015ida,Alonso:2016gym,Gonzalez-Alonso:2016etj}. Therefore, normalized observables together with
$R_{D^{(*)}}$ can discriminate between the two NP scenarios.  

To illustrate our argument let us start by fitting the tensor contribution to $R_{D^{*}}^{\rm expt}$ together with the experimental average of $R_D$~\cite{Amhis:2014hma},  
\begin{equation}
R_D^{\rm expt}=0.397(40)(28).
\end{equation} 
Adding the statistical and systematic error of the experimental results in quadratures and taking into account their relative negative 
correlation, $\rho=-0.21$~\cite{Amhis:2014hma}, we find the solution
\begin{equation}
\eT=0.377(12). \label{eq:eT_RDs}
\end{equation}
with a $\chi^2=1.49$. In contrast, in  the left-handed scenario we find $\eL=0.13(3)$  at the  minimum $\chi^2=0.013$.

One example of the normalized observables defined above is the longitudinal polarization of the $\tau$ in the decay, which is defined as~\cite{Tanaka:1994ay},
\begin{equation}
d P_L^{(*)}=\frac{d\Gamma_+-d\Gamma_-}{d\Gamma_++d\Gamma_-},\label{eq:LongPoltau_def} 
\end{equation}
where $d\Gamma_\pm$ denotes generically a differential decay rate (e.g. in $q^2$) of $B\to D^{(*)}\tau\nu$ and with 
the helicity of the $\tau$ being $\lambda_\tau=\pm$. The longitudinal polarization can be measured using the kinematic distributions
of the detected decay products of the $\tau$ as polarimeters~\cite{Bullock:1992yt,Tanaka:2010se} and a very first measurement of the 
integrated longitudinal polarization for the $BD^*$ channel, $P_L^*$, has been reported using the $\tau\to\pi\nu$ and $\tau\to\rho\nu$ modes in the full Belle 
dataset~\cite{Abdesselam:2016xqt}, 
\begin{equation}
P_L^{*,{\rm expt}}=-0.44(47)^{+0.20}_{-0.17},\label{eq:PLs_expt} 
\end{equation}
which is consistent with the SM prediction, $P_L^{*,{\rm SM}}=-0.504(24)$. Using the result of the fit obtained in Eq.~(\ref{eq:eT_RDs}), 
we obtain, 
\begin{equation}
P_L^{*}=0.190(10).\label{eq:PLs_tens}  
\end{equation}
This prediction of the tensor scenario leads to a $P_L^*$ with the opposite sign compared to the SM and stands at about $1\sigma$ from the experimental measurement 
in Eq.~(\ref{eq:PLs_expt}). To be more quantitative, we can add 
$P_L^*$ to the fit of $\eT$ to the experimental averages of $R_{D^{(*)}}$ performed above. Assuming no further correlation an adding the errors in Eq.~(\ref{eq:PLs_expt})
in quadratures,   
the position of the minima in Eq.~(\ref{eq:eT_RDs}) does not change and the $\chi^2$ increases to $\sim3$. This suggests that the tensor scenario is disfavored 
with respect to the left-handed one, although more precise measurements of $P_L^*$ (or other normalized observables) would be essential to confirm or refute 
this conclusion with a higher statistic significance. This is illustrated on the right-hand panel of Fig.~\ref{fig:BctaunuRDs}, where we display the
correlation between $R_{D^*}$ and $P_L^*$ as parametric functions of $\eT$ compared to the current experimental measurements. We also show  prospects
estimated naively assuming an order-of-magnitude increase in statistics and improving the systematics by a factor of two.    

We end the discussion by pointing out a caveat in our latter conclusions. The tension between the ``tensor interpretation'' of the anomalies and data can 
be relaxed if one adds the coefficient $\eS$, which exclusively contributes to the $B\to D\tau\nu$ channel, to the fit. One can, indeed, see that 
in this case $R_{D^*}$ can be explained by a small negative value of $\eT$ that interferes constructively with the SM in the $B\to D^*\tau\nu$ channel and does not contribute
to $P_L^*$ significantly. This solution interferes destructively in the  $B\to D\tau\nu$ channel, producing a deficit in $R_D$ that needs a moderate 
value of $\eS$ to make it consistent with the experimental value. The effect of this combined $\eT$-$\eS$ scenario is represented by the dashed-line 
that developes to the right of the SM in the right panel of Fig.~\ref{fig:BctaunuRDs}. It can be distinguished from the other scenarios above efficiently
only by looking for the effects of $\eS$ in normalized observables of the  $B\to D\tau\nu$ channel such as $P_L$.

\section{Conclusions}

We have discussed a powerful constraint on possible new-physics interpretations of the $R_{D^{(*)}}$ anomalies that
is derived from the lifetime of the $B_c$ meson. This is obtained by demanding that the contribution rate of the $B_c\to\tau \nu$ decay channel does not 
exceed the fraction of the total width that is allowed by the calculation of the lifetime in the Standard Model. This leads to a very strong bound on 
scenarios manifest at low energies as 4-fermion scalar contact operators, since they lift the slight (but not negligible) chiral suppression of the 
$B_c^-\to\tau^-\bar\nu$ amplitude in the Standard Model. The new constraint is sufficient to rule out, model-independently, explanations of the 
enhancement in $R_{D^*}$ based on scalar operators because the $B\to D^*\tau \nu$ decay amplitude depends on exactly the same pseudoscalar combination 
of the corresponding Wilson coefficients as the one contributing to the pure tauonic $B_c$ decay. Popular new-physics models implementing extra 
Higgs doublets or certain classes of leptoquarks enter into this category. Updated calculations of the $B_c$ lifetime in QCD and measurements of the branching fractions of its decay channels at the LHC should refine the bound in the future.

Taking into account the constraint of the lifetime of the $B_c$ meson simplifies the new-physics analyses of the $R_{D^{(*)}}$ anomaly. For instance, 
only two effective operators, left-handed or tensorial, remain at low energies to account for the enhancement in $R_{D^*}$. We argued that the precise 
measurement of normalized observables, such as the longitudinal polarizations, could become instrumental to distinguish between the different scenarios.   

\section{Acknowledgments}
This work was supported in part  by the US Department of Energy under grant DE-SC0009919.

\bibliographystyle{kp}
\bibliography{BtoDtaunu.bib}
\end{document}